\documentclass{webofc}
\usepackage[varg]{txfonts}   
\usepackage{multicol}
\usepackage{adjustbox}
\usepackage{array}
\usepackage{caption}
\usepackage{subcaption}
\newcolumntype{M}[1]{>{\centering\arraybackslash}m{#1}}
\begin{document}
\title{Ranking-based neural network for ambiguity resolution in ACTS}
%
%

\author{\firstname{Corentin} \lastname{Allaire}\inst{1}\fnsep\thanks{\email{corentin.allaire@cern.ch}} \and
        \firstname{Françoise} \lastname{Bouvet}\inst{1} \and
        \firstname{Hadrien} \lastname{Grasland}\inst{1} \and
        \firstname{David} \lastname{Rousseau}\inst{1}
}

\institute{Universit\'e Paris-Saclay, CNRS/IN2P3, IJCLab, 91405 Orsay, France}

\abstract{%
The reconstruction of particle trajectories is a key challenge of particle physics experiments, as it directly impacts particle identification and physics performances while also representing one of the main CPU consumers of many high-energy physics experiments. As the luminosity of particle colliders increases, this reconstruction will become more challenging and resource-intensive. New algorithms are thus needed to address these challenges efficiently. One potential step of track reconstruction is ambiguity resolution. In this step, performed at the end of the tracking chain, we select which tracks candidates should be kept and which must be discarded. The speed of this algorithm is directly driven by the number of track candidates, which can be reduced at the cost of some physics performance. Since this problem is fundamentally an issue of comparison and classification, we propose to use a machine learning-based approach to the Ambiguity Resolution. Using a shared-hits-based clustering algorithm, we can efficiently determine which candidates belong to the same truth particle. Afterwards, we can apply a Neural Network (NN) to compare those tracks and decide which ones are duplicates and which ones should be kept. This approach is implemented within A Common Tracking Software (ACTS) framework and tested on the Open Data Detector (ODD), a realistic virtual detector similar to a future ATLAS one. This new approach was shown to be 15 times faster than the default ACTS algorithm while removing 32 times more duplicates down to less than one duplicated track per event.
 }
\maketitle
\section*{Introduction}
\label{intro}
\begin{figure}[htbp]
  \center
  \includegraphics[width=0.90\textwidth]{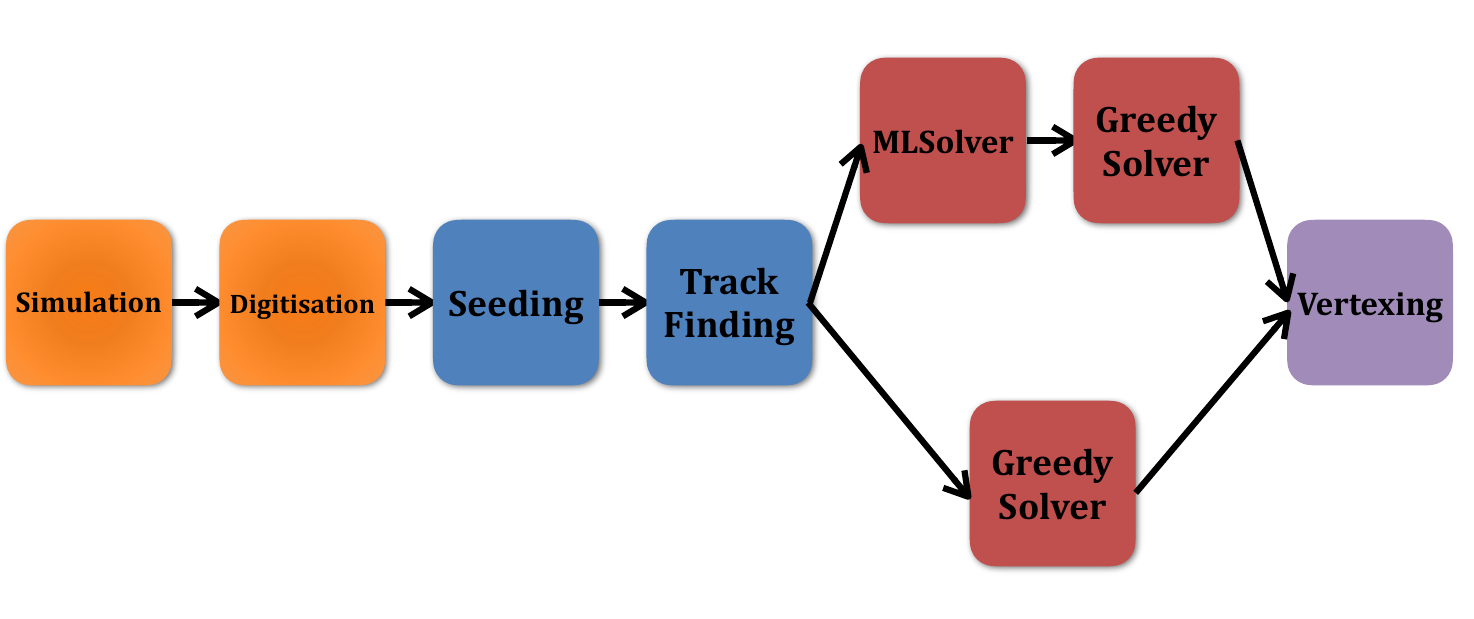}
  \caption{Illustration of the ACTS tracking chain with both Ambiguity Resolution options.}
  \label{Fig:full_chain}
\end{figure}

	Charged particle trajectory reconstruction is essential to most high-energy physics experiments. A precise measurement of those trajectories allows us to properly reconstruct the primary interaction vertex, identify the particles and evaluate their momentum. Trajectory reconstruction is also one of the most computationally intensive components of event reconstruction, as it scales quadratically with the number of particles in the detector. In this context, great efforts are being deployed to optimise such algorithms' performances in terms of physics object reconstruction and computing efficiency. One of these algorithms that could be improved is the ambiguity resolution. This corresponds to the last step of the tracking chain illustrated in Figure \ref{Fig:full_chain}. After reconstructing all the possible trajectories, it removes the fake ones (coming from an arbitrary combination of hits) and the duplicated ones (which have been reconstructed multiple times). The speed of this algorithm will tend to scale with the number of particles in the detector; it can thus become relatively slow in a high pile-up environment such as the one expected at the HL-LHC\cite{Apollinari:2284929}.
	
	This paper proposes a machine learning-based solution for ambiguity resolution. This new algorithm works in two steps. First, we cluster all trajectories assumed to come from the same truth particle. This is achieved using a shared-hits-based clustering algorithm. We can then use a neural network to score each reconstructed trajectory. The highest score in each cluster is then assumed to be the track that best matches the corresponding truth particle. 

	The use of machine learning (ML) for this algorithm should offer two advantages. Firstly, by learning from the trajectory in the detector, our new algorithm should easily adapt to different experiments and properly account for their inhomogeneity. Secondly, by replacing the complex original algorithm with what is essentially a series of simple matrix multiplication, we should be able to increase the speed of the algorithm, even with equivalent hardware.

\section{ACTS and dataset}
\label{Acts}

	A Common Tracking Software (ACTS)\cite{Ai:2021ghi} is a tracking framework being developed since 2016 as an international collaboration with the goal of providing a generic, experiment-independent open-source software framework for charged particle track reconstruction. The ACTS framework implements a virtual detector for testing purposes, the Open Data Detector (ODD)\cite{corentin_allaire_2022_6445359}. It is a typical all-silicon LHC tracking detector with ten layers of cylinders and disks, also including all the support structure and cabling of a realistic detector, and is implemented using DD4Hep\cite{Frank_2014}. It has been used as a reference for many developments in track reconstruction in ACTS. It can thus be used to test the performances of machine learning-based tracking algorithms and compare them with their classical equivalent\cite{Amrouche2023}.  

	Our algorithm has been fully implemented in ACTS and can thus be easily used by any ACTS-based experiments. It comes bundled with a trained network for the ODD and can be used with the full track reconstruction chain of the ODD by simply changing one option. Its integration in ACTS also means that any experiment which has implemented its detector in ACTS can use it. All they need to do is retrain the network for their detector with the script provided in the ACTS GitHub repository and then replace the ambiguity resolution algorithm in their reconstruction chain.  

	The training of our network and the evaluation of its performances have been done using simulated $t\bar{t}$ events (which are the standard for tracking performance evaluation) under conditions similar to the HL-LHC: an energy in the centre of mass of $\sqrt{s}=14$ TeV and 200 additional pile-up vertices per event. The events were generated using Pythia8\cite{10.21468/SciPostPhysCodeb.8}, and only particles with a transverse momentum ($p_{T}$) of more than 1~GeV were considered. In total, 2000 events were generated; 1000 were used in the training of the network, while the remaining 1000 were used in testing our performances. For those events' signatures, 1000 events correspond to 12M tracks reconstructed by ACTS and 800k truth particles in the detector. Further studies on the impact of the training sample size are planned.

\section{Ambiguity resolution and ACTS Greedy Solver}
\label{Greedy}

	ACTS implements, by default, an ambiguity solver algorithm called the Greedy Solver. As its name implies, this algorithm takes a series of locally optimal choices for trajectory removal. This is not guaranteed to yield the best trajectory representation of the particle but should offer a decent selection relatively quickly. 
	
	The ambiguity resolution with this algorithm is performed by first choosing the maximum number of shared hits a track can have, with a shared hit defined as a hit belonging to multiple trajectories. We decided to have at most two shared hits for the ODD with ACTS, but this can depend on the detector and the expected physics performances (small values lead to high purity but small efficiency). This criterion is used since we expect hits to only belong to one trajectory. If two tracks share a hit, one of them is usually incorrect. In some cases, hits are expected to belong to multiple trajectories (for example, in the case of a merged cluster), with most experiments having algorithms to recognise those cases\cite{Aad:1712337}. Currently, ACTS does not handle this, but this is being implemented. 

	The first step of the algorithm is to select all the trajectories not matching this criterion, in the ODD case, three or more shared hits. Among those tracks, we identify the one with the largest fraction of shared hits. Since a track with mostly shared hits is either redundant or fake, we remove it. If there is a tie, the track with the largest $\chi^{2}$ is removed. Afterwards, we recompute the number of shared hits in all the remaining tracks and repeat the previous step until all the remaining tracks have less than the maximum number of shared hits. Those remaining tracks are thus considered good and can be passed to the following algorithm in our tracking chain.

	This algorithm has been shown to perform well but does have some drawbacks. Firstly, it can get relatively slow when the number of tracks is very large, as we need to recompute the number of shared hits after each track is removed. Secondly, since it uses little track information, we are not guaranteed to select the best track representation of the initial particle, which could slightly reduce physics performance.

\section{Ambiguity Solving with a Ranking-Based Neural Network}
\label{MLSolver}

	Our Ambiguity Solver with a Ranking-Based Neural Network, called MLSolver in ACTS, proposes a different approach to Ambiguity resolution. Instead of iteratively removing the worst track in the detector until all the tracks pass a specific quality criteria, it tries to cluster together all tracks that appear to come from the same particle and then select the best one within this cluster. To do this, the algorithm is split into two steps: first, the clustering of the track using hit-sharing information followed by a neural network-based track scoring to decide which one should be kept.
	
	The clustering aims to create at least one cluster per original truth particle. Having more clusters than particles is preferable to having less, as the second case will result in a loss of efficiency. To start the clustering, we hypothesise that the more hits a track has, the better it usually is. Our clusters are built around a leading track with a high number of hits; all the other cluster tracks are then track-sharing hits with that cluster's leading track. 
	
	To create the clusters, the algorithm loops through all the tracks in the detector, starting with the one with the most hits and ending with the one with the least. For each track, it checks if its hits are shared with the leading track of one of the clusters. If yes, that track is added to that cluster. If not, a new cluster is created with the current track as its leading track. This process is then repeated until all the tracks belong to a cluster. Note that the hits-sharing comparison is only performed with the leading track and not the other tracks in the clusters.
	
	The clustering step directly affects the track reconstruction efficiency and the duplicate rate. Thus, for complex or unusual particle detectors, the hit-sharing-based clustering can easily be replaced by a more complex algorithm more adapted to the experimental condition.
	
	The next step after the clustering is determining which track corresponds to the original particles in each cluster. To do this, we use a four-layer neural network that is going to score the different trajectories based on different track information: 
\begin{multicols}{2}
\begin{itemize}
	\item Number of states
	\item Number of measurements
	\item Number of outliers measurements
	\item Number of holes
\end{itemize}
\columnbreak
\begin{itemize}
	\item Number of degrees of freedom (NDF)
	\item $\chi^{2}/NDF$
	\item Direction in $\eta$
	\item Direction in $\phi$
\end{itemize}
\end{multicols}  
The network uses three fully connected hidden layers with 10, 15 and 10 neurones each, all using a ReLU activation function. The output layer, on the other hand, uses a sigmoid activation function to compute a score between 0 and 1. With those variables, the network will compute a score for each trajectory, ideally reflective of the quality of the track. The algorithm can then select the track with the highest score for each cluster and remove the others. Fundamentally, this network aims to rank the tracks in the different clusters from the most likely to be the correct one to the least likely. To achieve this, we train our network using a margin ranking loss function.

	To train a neural network with ranking capability, we compute one loss per cluster in the input, which will evaluate how good the network is at separating the good track from the fake and duplicate. In training, the clustering efficiency is assumed to be $100\%$ efficient. In that case, we can directly use the truth information to perform the clustering; each cluster will be composed of all the tracks associated with the same truth particle. This truth association is performed by looking at the truth particle with the most hits in common with each track. For each cluster, we can then determine which track is good. It is defined as the track with the most shared hits with the truth particle, then the least outlier and finally the smallest $\chi^{2}$.

	The loss function being minimised for the different clusters is the margin ranking loss, a loss function commonly used for information retrieval and metric learning, shown in Eq\ref{MRL}. In this equation, x is the score of a track in the cluster, y is the score of the good track of the cluster (there is one since we use truth clustering), and margin is a user-defined margin between the good score and the others. Ultimately, by minimising this function, the network will learn to separate the good and bad tracks in each cluster.
	
\begin{equation}
Loss_{cluster} = \frac{1}{N_{tracks}}\sum^{tracks}max(0, x - y + margin)
\label{MRL}
\end{equation}
	
	One of the main advantages of this technique is that since it only uses the relative value of scores within the different clusters, it can automatically account for inhomogeneity in the detector since it doesn't try to identify which is the best track, only the best one in a small constraint region. Finally, it is also orthogonal to the approach used by the Greedy Solver; this means that if the MLSolver does not remove some duplicates, we can call the Greedy Solver on the remaining tracks to remove them. Since only a few tracks are still present after the MLSolver, this will barely affect the algorithm's speed while further improving the performance. This approach will be used in Section \ref{res}.
	
\begin{table}
\centering
\caption{Performance of the Greedy and MLSolver on $1000$ $t\bar{t}$ events}
\label{tab}       
\begin{adjustbox}{width=\textwidth}
\begin{tabular}{|M{1cm}|M{1.7cm}|M{1.9cm}|M{1.9cm}|M{1.9cm}|M{1.5cm}|M{1.5cm}|M{1.5cm}|}
\hline
                                   & Number of tracks & Number of truth particles & Efficiency (good tracks) & Efficiency (truth tracks) & Duplicate Rate & Fake Rate & Solver speed [ms/event]  \\\hline
CKF                            & 12556  & 827.3 & 100 \%  & 100 \% & 93.0 \% & 0.39 \% & 0 \\\hline
CKF + Greedy Solver Only & 842.2 & 826.4 & 79.5 \% & 99.9 \% & 0.64 \% & 1.24 \% & 398 \\\hline
CKF + MLSolver        & 823.8 & 818.9 & 87.2 \% & 99.0 \% & 0.02 \% & 0.57 \% & 26.6 \\\hline
\end{tabular}
\end{adjustbox}
\end{table}	
	
\section{Results}
\label{res}

	The Machine learning-based ambiguity solving has been fully implemented in ACTS and can easily be used to replace the greedy solver in the full tracking chain. The algorithm uses Onnxruntime \cite{onnxruntime} to perform the inference inside the ACTS C++ code, importing the network via an Onnx file. For the ODD, such a file is available with the ACTS Examples by default, allowing everybody to test it quickly. For other detectors, the training of the network is easy to perform. By running the ACTS full tracking chain, one can extract the different trajectories as a CSV file. This file can then be used as input to a provided Python script to train the network and obtain the Onnx file.

	The performance and efficiency of this algorithm have been tested using simulated $t\bar{t}$ events in the ODD. We use the full tracking chain provided in the ACTS Examples up to the track reconstruction step using the Combinatorial Kalman Filter (CKF) to obtain the trajectories in the detector. For the training and the performance study, we only use tracks with more than seven hits as we consider tracks with less than seven hits unusable for the ODD.
	
	To study the algorithm's performance, we first must associate all our trajectories with a truth particle. This is done by associating each track with the truth particle with which it shares the most hits. Afterwards, we can define three types of tracks after the CKF : 
\begin{itemize}
	\item Good tracks: for a given truth particle, most truth match measurements, then fewer outliers, then the smallest $\chi^{2}$.
	\item Duplicated tracks: more than $50\%$ truth-matched measurements (but is not the good track). 
	\item Fake tracks: less than $50\%$ truth matched measurements. 
\end{itemize}

	With those three categories of track defined, we can study the physics performances of the algorithms. Ideally, after the solver, we should have only one track per truth particle after ambiguity resolution, and this track should be a good track and not a duplicate. This is evaluated by looking at the following variables :
\begin{itemize}
	\item Truth particle efficiency: fraction of truth particles with an associated track before the ambiguity solving that still have one after. It is related to the clustering quality.
	\item Good track efficiency: fraction of good track still present after the ambiguity solving. It is related to the network quality.
	\item Duplicate rate: fraction of duplicate track. It is related to the clustering quality.	
	\item Fake rate: fraction of fake track. It is related to the clustering and network quality.
\end{itemize}
		
	In addition to those, we will also look at the speed of the algorithms on the same CPU in ms/events. Those measurements have been performed on a local laptop, informing us on the relative speed of both methods.
	
	The performance of both the Greedy Solver and the ML Solver are shown in Table \ref{tab}. Both algorithms exhibit a high truth particle efficiency.

\begin{figure}[htbp]
\centering
	\begin{subfigure}{0.48\textwidth}
		\includegraphics[width=\textwidth]{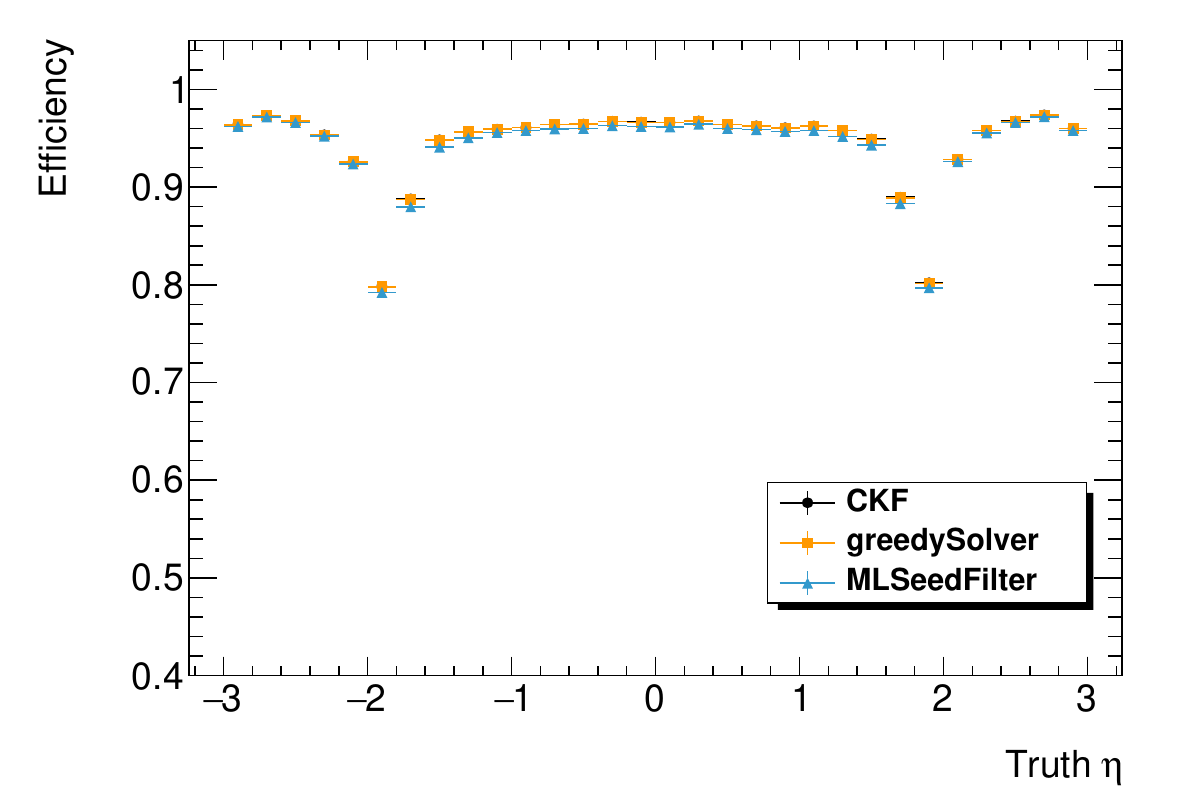}
		\caption{}
	\end{subfigure}
	\begin{subfigure}{0.48\textwidth}
		\includegraphics[width=\textwidth]{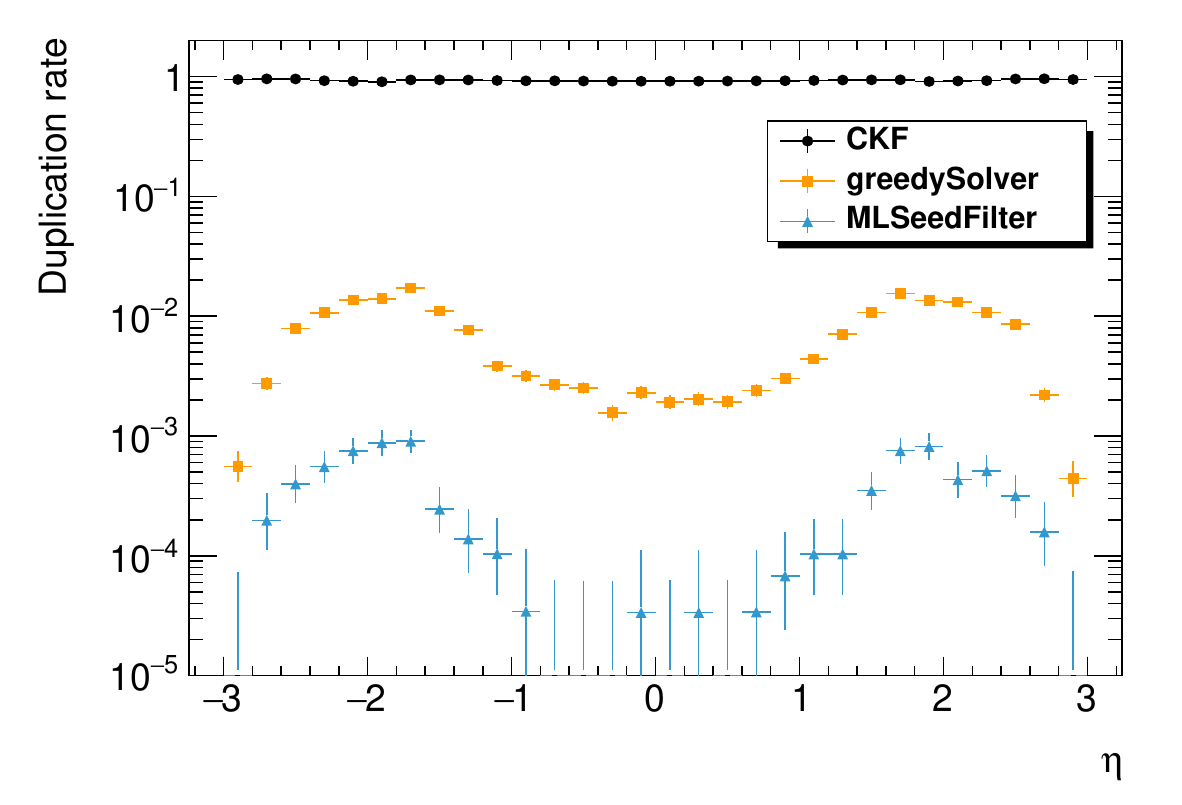}
		\caption{}
	\end{subfigure}
	\newline
	\begin{subfigure}{0.48\textwidth}
		\includegraphics[width=\textwidth]{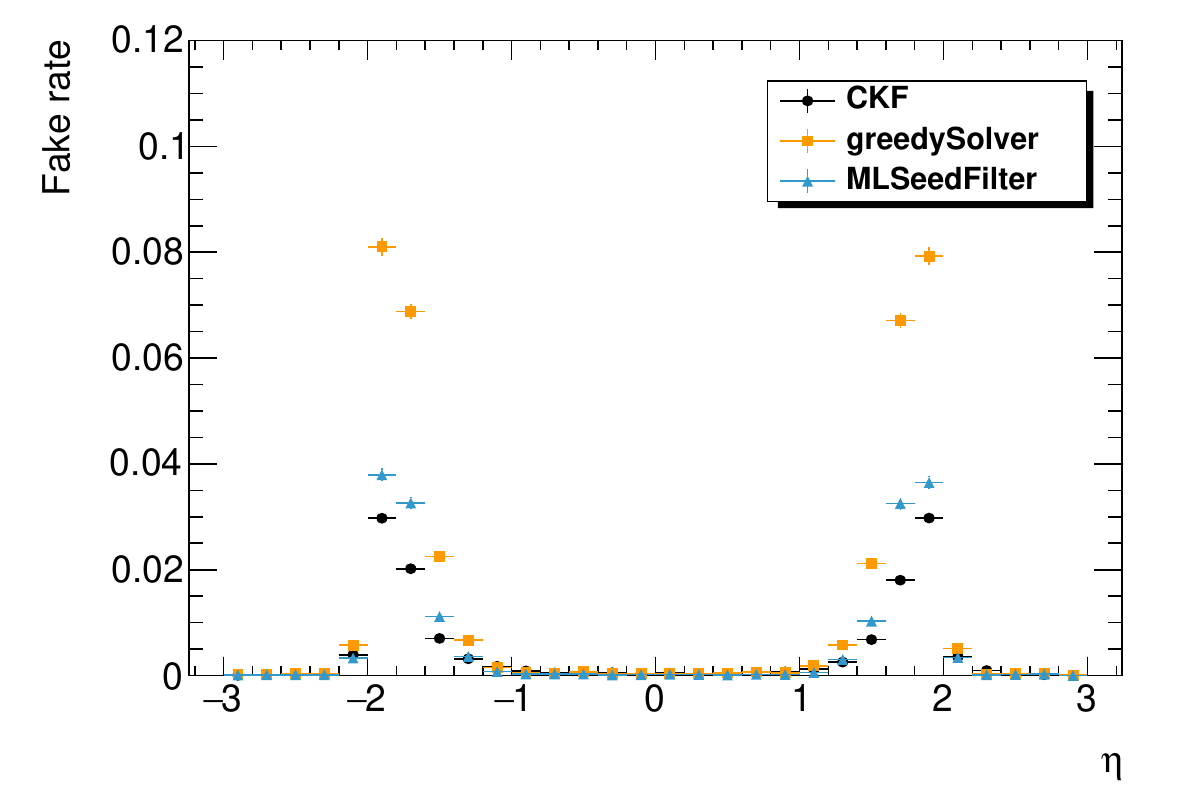}
		\caption{}
	\end{subfigure}
	\caption{(a) Track efficiency, (b) duplicate rate and (c) fake rate shown as a function of the pseudo-rapidity $\eta$ in the detector for the CKF alone, the Greedy Solver and the MLSolver.}
  \label{Fig:perf_eta}
\end{figure}
		
		The truth particle efficiency is the highest for the Greedy Solver with $99.9\%$, while the MLSolver is a bit lower with only  $99.0\%$, representing a loss of roughly eight particles per event. That loss can be acceptable, but further studies are needed to identify precisely the impact on particle reconstruction. On the other hand, the MLSolver significantly increases the good track efficiency from $79.5\%$ to $87.2\%$; this means that even if we reconstruct fewer trajectories with this approach, they are of greater quality and are closer to the original truth particles. The impact of this efficiency improvement on the reconstruction performance will be part of a future study using the more realistic detector. Finally, the new ML approach improves on both types of inefficiency represented by the Duplicate and Fake Rate. The Fake Rate decreases by a factor of 2.2, and the Duplicate Rate decreases by a factor of 32 to, on average, less than one remaining duplicated track per event. This means we can be much more confident about the origin of the remaining tracks after the MLSolver. In addition to the improvement in physics performances, the other significant improvement of the ML algorithm is in speed of execution. For the considered events, the MLSolver is 15 times faster than the Greedy one. This improvement should scale with the number of tracks in the events as the ML algorithm is expected to scale linearly with the number of tracks, while the Greedy one should scale quadratically. 
	
\begin{figure}[htbp]
\centering
	\begin{subfigure}{0.48\textwidth}
		\includegraphics[width=\textwidth]{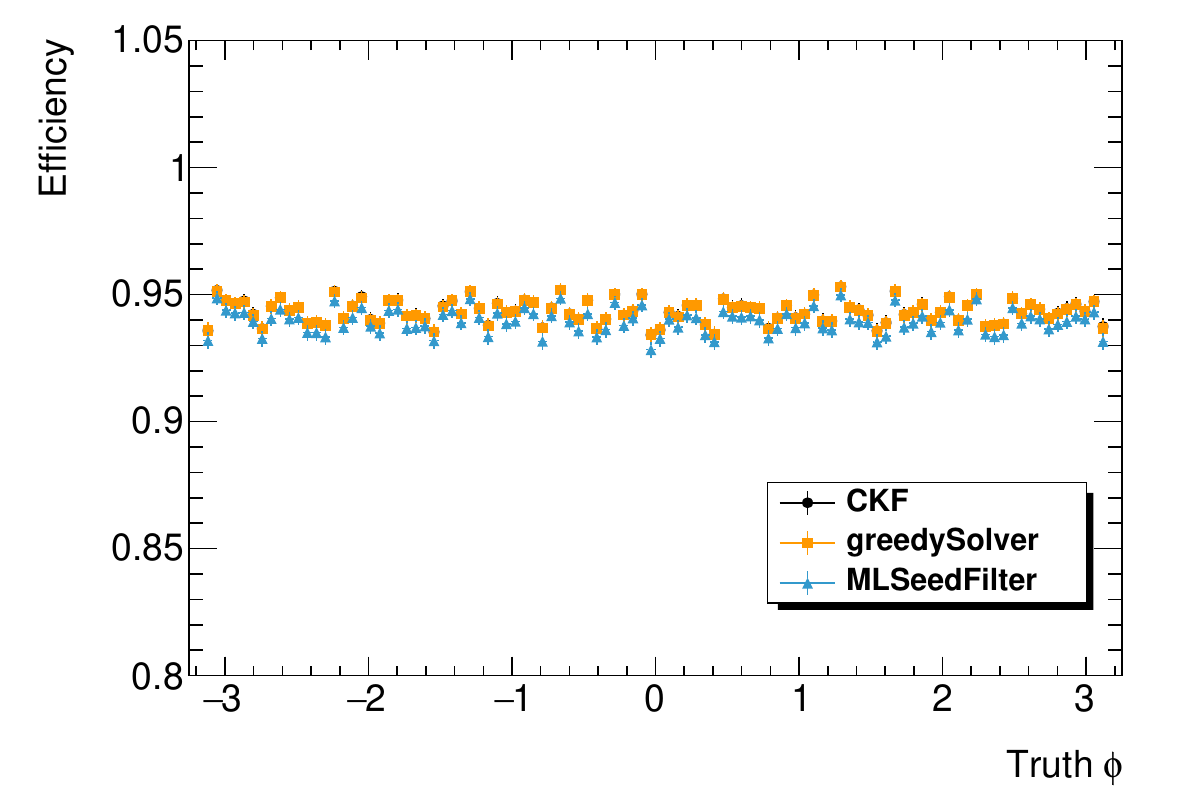}
		\caption{}
	\end{subfigure}
	\begin{subfigure}{0.48\textwidth}
		\includegraphics[width=\textwidth]{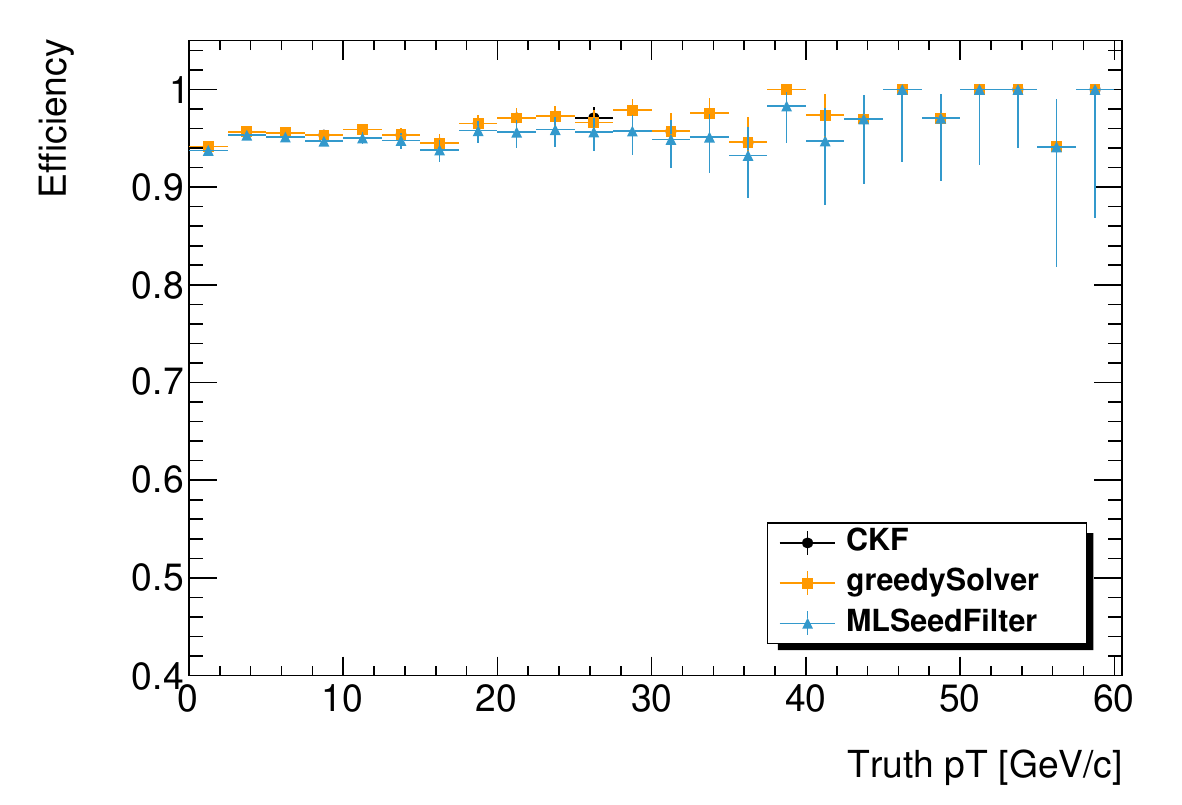}
		\caption{}
	\end{subfigure}
	\newline
	\begin{subfigure}{0.48\textwidth}
		\includegraphics[width=\textwidth]{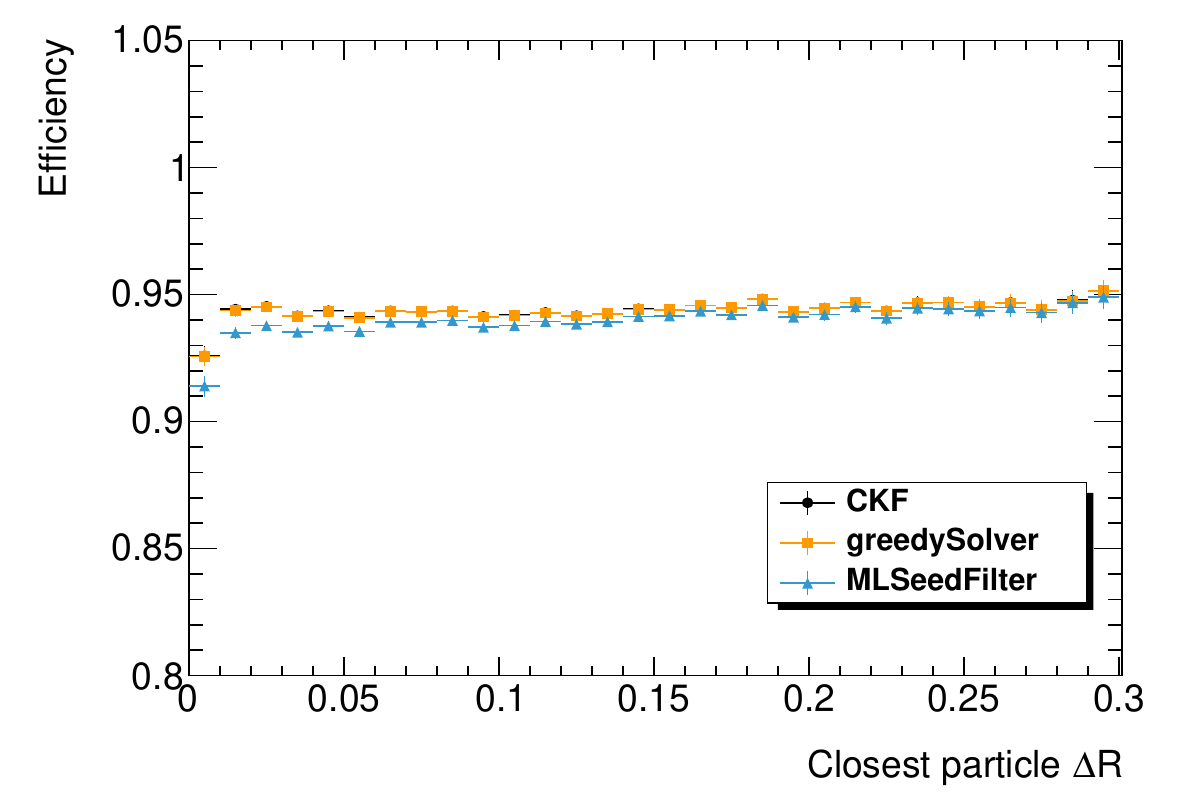}
		\caption{}
	\end{subfigure}
	\caption{Track efficiency as a function of (a) $\phi$, of the (b) transverse momentum $p_{T}$ and of (c) the distance to the closest other truth particle. Those are shown for the CKF alone, the Greedy Solver and the MLSolver.}
  \label{Fig:perf_eff}
\end{figure}

	In addition to global performance for the entire detector, we have also looked at the performance of the Solver algorithms in the different regions of the detector. This is illustrated in Figures \ref{Fig:perf_eta}, which shows the efficiency, fake rate and duplicate rate as a function of the pseudorapidity $\eta$. Loss of efficiency is observed around $\eta = 2$. Still, they are expected as they correspond to the transition between the central barrel sensor and forward endcap one, leading to a less instrumented region. Similar shapes are seen for the two different algorithms, confirming that the use of machine learning doesn't induce any additional inhomogeneity in the efficiency of the reconstruction. Similar studies have been performed as a function of $\phi$ and of the particle's transverse momentum, and a similar homogeneity has been observed as shown in Figure \ref{Fig:perf_eff}. Finally, the efficiency is shown as a function of the distance in $\Delta R$ to the closest different truth particle in Figure \ref{Fig:perf_eff} (c). We can see that our new algorithm has its worst efficiency in dense environments such as jets, which is expected due to the high number of overlapping hits in this region, but even in those regions, the efficiency loss compared to the CKF is relatively small only being of the order of $1\%$.

\newpage

\section*{Conclusion}
\label{conclusion}

	We have shown that ambiguity resolution can be significantly improved by using a neural network with a margin ranking loss function. Both in terms of physics performance with a reduction of the duplicate rate by a factor of 32 down to 0.2 duplicated track per event, and speed of execution with a speed-up of a factor of 23. The new MLSolver is fully available in ACTS and is pre-trained for use with the Open Data Detector; all the scripts for training and performance evaluation with any other detector implemented in ACTS are also provided in the ACTS GitHub repository. The next steps for this project are the use of this algorithm with a more realistic detector (such as the ATLAS ITk \cite{CERN-LHCC-2017-021}\cite{CERN-LHCC-2017-005}) and in more complex tracking environments such as dense jets.

\section*{Acknowledgements}
This project has received funding from the European Union’s Horizon 2020 research and innovation programme under grant agreement No 101004761. \newline

\bibliography{Bibliographie}

\end{document}